\documentstyle[preprint,aps,epsf,floats]{revtex} 

\def\GeV{\ \rm{ GeV} }

\begin{document}

\preprint{\tighten \vbox{\hbox{CMU-HEP 98-03} }}

\title{$J/\psi$ Production at LEP: Revisited and Resummed}

\author{C.\ Glenn\ Boyd, Adam K.\ Leibovich, and I.\ Z.\ Rothstein}

\address{
Department of Physics,
Carnegie Mellon University,
Pittsburgh, PA 15213}

\maketitle

{\tighten
\begin{abstract}

We present the leading order differential and total 
rates for $J/\psi$ production at LEP.
By leading order we mean all terms of the form $\alpha_s[\alpha_s 
\log(M_Z^2/M_{\psi}^2)]^n$ and $\alpha_s^{n+1} \log^l(z^2) 
\log^m(M_Z^2/M_{\psi}^2),~ (l+m=2n-1)$, in the regions $z=2E_\psi/M_Z
\sim O(1)$ and $z \ll 1$, respectively. In the intermediate region we
interpolate using the available data.  This resummation eliminates the
$O[\alpha_s(M_\psi)/\alpha_s(M_Z)]\sim 2$ theoretical uncertainties in
previous calculations.  The $\log(z)$ resummation results in a
suppression of the small $z$ region due to coherent gluon emission.
Comparing the zeroth moment with the LEP data we find the value for
the effective octet matrix element to be $\langle \hat
O_8^{\psi}(^3S_1)\rangle=0.019\GeV^3$.  The theoretical uncertainties
are substantially smaller than those from Tevatron extractions.  Using
this value of the octet matrix element we make a prediction for the
first moment of the differential rate and find that the resummed
differential decay rate is in much better agreement with preliminary
data than the color singlet result or the unresummed color octet
prediction.

\end{abstract}
}%end tighten

%%\pacs{}

\newpage 

\section{Introduction}

The subject of quarkonium production has gained renewed interest due
to the fact that it can now be understood from first principles.  The
production rates are calculated in a systematic expansion in
$\alpha_s$ and $v$, the relative velocity of the heavy quarks in the
rest frame of the quarkonium bound state.  This is accomplished by
working in non-relativistic QCD (NRQCD) where the expansion in $v$ is
implemented by utilizing the scaling properties of non-perturbative
matrix elements\cite{BBL}. The use of this effective theory has
clarified formal issues and allowed for better fits to data.  In
general, NRQCD leads to larger cross sections than its historical
predecessor, the color singlet model, largely because NRQCD predicts
substantial contributions to various cross sections from the color
octet channel.  This happens even though the octet matrix elements are
suppressed by powers of $v$, because the color singlet channels can
themselves be suppressed either kinematically or by powers of
$\alpha_s$.  For instance, $\psi'$ production at the Tevatron can now
be well fit if one allows for color octet production\cite{BratFlem}.

The field of quarkonium production itself has  matured to the point
where we would like to go beyond order of magnitude accuracy.  Indeed, 
once the formalism is verified quantitatively, it can be used as
a tool in other areas in strong interaction physics, such as heavy 
ion collisions and the measurement of the spin dependent gluon
distribution functions. 

Presently, the values for the octet matrix element $\langle
O_8^\psi(^3S_1) \rangle$ and a certain linear combination of the
$\langle O_8^\psi(^3P_0) \rangle$, and $\langle O_8^\psi(^1S_0)
\rangle$ matrix elements have been extracted at the Tevatron. These
matrix elements should be universal in nature, in that we should be
able to use them as input in other processes to make definite
predictions.  Unfortunately, hadronic uncertainties in the Tevatron
extraction make this difficult.  The extraction depends sensitively on
the choice of gluon distribution function and factorization scale, as
well as how one treats initial gluon radiation\cite{Ocl}. Extractions of
$\langle O_8^\psi(^3S_1) \rangle$ are as disparate as $2.1 \times
10^{-3}$ \cite{Ocl}, $2.7 \times 10^{-3}$ \cite{Okk}, $6.6 \times
10^{-3}$ \cite{ChoLeib}, and $14.0 \times 10^{-3}$ \cite{Obk},
in$\GeV^3$. A process involving smaller theoretical uncertainties is
needed. Perhaps the cleanest setting to extract a value of the octet
matrix element is in prompt $J/\psi$ production at LEP, because in
lepton initiated processes the theoretical errors are bounded by our
computational strength rather than higher twist effects.  Furthermore, 
at LEP we can measure the energy distribution as well, thus once
we've extracted the value of the octet matrix element, we can then
make predictions for the moments of the rate.  This provides a strong test for
the color octet mechanism.

Formally there are two leading order contributions in the $\alpha_s$ and $v$
expansion, both in the singlet channel, which are of order
$O(\alpha_s^2 v^3)$.  There is a contribution from gluon radiation in
the singlet channel $Z \to \psi g g$ that is suppressed by powers of
$M_\psi^2/E_\psi^2$\cite{KS}. There is also the color singlet charm
quark fragmentation process $Z \to\psi c {\overline c}$\cite{BCK,BCY},
which has no power suppression and thus dominates over
non-fragmentation processes, for large $E_\psi$.  Light quark octet
fragmentation (in which the mother parton does not combine to form
part of the bound state) is naively of order $\alpha_s^2 v^7$, down by
$v^4 \sim 1/10$ compared to charm fragmentation.  However, as it turns
out, this channel is enhanced due to the presence of large logs as
well as a numerical factor of five due to the number of possible
quarks that initiate the process.  Indeed, previous calculations of the
$J/\psi$ production rate at LEP are dominated by these light quark
fragmentation contributions \cite{Cho,CKY}. They give cross sections
that are of the correct order of magnitude when values of the
non-perturbative matrix elements are taken from the Tevatron fits
\cite{ChoLeib}.  However, the same logs that enhance the octet channel
also put the convergence of the perturbative expansion into question.

The tree-level calculation of the differential cross section in the
color octet production channel\cite{Cho,CKY,YQC} is enhanced by a
large logarithm, ${d\Gamma\over{dz}}(Z \to \psi +X) \sim
\alpha_s^2 \log(M_Z^2/M^2_\psi)/z$, leading to large double logs in the
total rate. Since $\alpha_s \log(M_Z^2/M^2_\psi)
\approx 1.5$, we should treat $\alpha_s \log(M_Z^2/M^2_\psi)$ as order one
and resum all powers of the large logarithm. With this counting, the
octet channel is $O(\alpha_s^0 v^7)$, on par with the singlet
fragmentation contribution.  More practically, the tree-level
calculation has a factor of two uncertainty associated with the scale
at which $\alpha_s$ is evaluated, since $\alpha_s(M_\psi)
/\alpha_s(M_Z) \approx 2$ (this is just a restatement that there is a
large logarithm).  The resummation of the leading logarithms
eliminates this uncertainty, so the resummation procedure is essential
from both a practical and a formal standpoint.  We therefore calculate
the quarkonium differential production rate at LEP taking all terms of
the form $\alpha_s^{m+1} \log^m(M_Z^2/M^2_\psi)$ as leading order.  We
will see that this resummation dramatically changes the differential
 cross section. However, summing the above mentioned logs
will only yield the correct leading order differential rate if $z$ is
sufficiently large.  When $z$ is parametrically small, terms of the
form $\alpha_s\log(z)/z$ become just as important. Furthermore, these
logs will also contribute double logs to the total rate given that the
lower limit on $z$ is $2 M_\psi/M_Z$.  This second type of log, due to
soft gluon emission, is resummed using a formalism familiar from
discussions of jet multiplicities\cite{rus,italy}.  Thus, we split
the calculation into two regimes, $z\sim1$ and $z \ll 1$.  We then
interpolate between these regimes using the data.

\section{Fragmentation Formalism}
The tree-level differential rate for color octet $J/\psi$ production 
is \cite{CKY}
\begin{eqnarray}\label{exact}
\frac{d\Gamma}{dz}(Z\to\psi(z)q\bar q) &=&
\frac{4\alpha_s^2}{9}\,\Gamma(Z\to q\bar q)\,\frac{\langle
O_8^\psi(^3S_1)\rangle}{M_\psi^3} \\ 
&& \times \left\{\left[\frac{(z-1)^2+1}{z} +
  2\frac{M_\psi^2}{M_Z^2}\frac{2-z}{z} +
  \frac{M_\psi^4}{M_Z^4}\frac{2}{z} \right]
  \log\left(\frac{z+z_L}{z-z_L}\right) - 2 z_L\right\},\nonumber
\end{eqnarray}
where the rescaled $\psi$ energy in the $Z$ rest frame 
$z=2E_\psi/M_Z$ has a physical range
of $2M_\psi/M_Z <z<1+M_\psi^2/M_Z^2$, 
$z_L=(z^2-4M_\psi^2/M_Z^2)^{1/2}$, and to the order we work, $M_\psi$ is
twice the charm mass. Performing the integration over
$z$ leads to the aforementioned double logs.
In the fragmentation limit, Eq.~(\ref{exact}) can be simplified to
\begin{equation}\label{fragRate}
\frac{d\Gamma}{dz}(Z\to\psi(z)q\bar q) \approx
\frac{4\alpha_s^2}{9}\Gamma(Z\to q\bar q)\frac{\langle
O_8^\psi(^3S_1)\rangle}{M_\psi^3} 
\left\{\frac{(z-1)^2+1}{z}\left[\log\left(\frac{M_Z^2}{M_\psi^2}\right)+
\log(z^2) \right] -
2z\right\}.
\end{equation}
In this limit, the differential rate
can be recast as the sum of quark and gluon fragmentation
processes,
\begin{equation}\label{fragForm}
\frac{d\Gamma}{dz}(Z\to\psi(z)\,q\bar q) = 2\, C_q(\mu^2,z) \ast
D_q(\mu^2,z) + C_g(\mu^2,z) \ast D_g(\mu^2,z).
\end{equation}
Here, $D_{q\to\psi}(z)$ and  $D_{g\to\psi}(z)$ are the 
light quark (or anti-quark) fragmentation 
and  gluon fragmentation functions respectively. The 
asterisk denotes convolution with the partonic production rates,
$ C \ast D \equiv \int_z^1 C(y) D(z/y) dy/y$ .
The $\mu$ dependence of the fragmentation functions is canceled by
that of the coefficient functions, $C_q$ and $C_g$.  All dependence on
$M_\psi$ is contained in the fragmentation functions, while all
dependence on $M_Z$ is contained in the coefficient functions. It is
this factorized form that will later allow us to resum the large
logarithms.

We choose to define the color octet fragmentation functions according 
to Collins and Soper\cite{CollinsSoper},
\begin{eqnarray}\label{fragfuncs}
D_{g\to\psi}(\mu^2, z) &=& \frac{-z^{d-3}}{16 (d-2) \pi k^+}\int dx^-
e^{-i P^+x^-/z} \langle 0 | G^{+ \nu}_b(0)\,a_\psi^\dagger (P^+,{\bf{0}})\,
a_\psi(P^+,{\bf{0}})\,G_{b,\nu}^+(0,x^-,{\bf{0}})| 0 \rangle,\nonumber\\
D_{q\to\psi}(\mu^2, z)& = &\frac{z^{d-3}}{4 \pi}\int dx^-
e^{-i P^+x^-/z} \frac{1}{3}{\rm Tr}_{\rm color}\,\frac12{\rm Tr}_{\rm Dirac}\\ 
&& \qquad \qquad \qquad \qquad \qquad\left[
\gamma^+ \langle 0 | Q(0)\,a_\psi^\dagger (P^+,{\bf{0}})\,
a_\psi(P^+,{\bf{0}})\,\bar{Q}(0,x^-,{\bf{0}}) | 0 \rangle \right]\,.\nonumber
\end{eqnarray}
Here, $G^{+\nu}_b$ is a gluon field strength tensor with color index 
$b$ and Lorentz indices $+$ and $ \nu$, $Q$ is a quark field, 
$a_\psi^\dagger$ is a creation operator for a $\psi$ meson, and $d$ is
the number of spacetime dimensions. The fragmentation functions 
are interpreted as the probabilities for a parton with momentum $k^+$
to decay into a $\psi$ with light cone momentum $P^+=z k^+$.
We have chosen to work in the light cone gauge, where eikonal
factors usually written to make gauge invariance manifest reduce to
the identity. An advantage of this definition over the alternative 
\cite{Fleming2} is its consistency with factorization at any 
subtraction scale $\mu$.

For the case of quarkonium fragmentation $D_{q\to\psi}$ and $D_{g\to\psi}$ 
can be calculated in a systematic expansion in $\alpha_s$ and
$v$ by matching onto NRQCD. Any soft divergences
which may arise due to the semi-inclusive nature of the
process cancel in the matching. Since we are working to leading order
no such divergences occur, and the matching is trivial in the sense
that there are no corrections to be calculated in the effective theory.
The calculation of these fragmentation functions are well documented
in the literature\cite{BY,MA}, so here we just present our results which
are in agreement with these previous calculations.

In the ${\overline {\rm MS}}$ scheme we find
\begin{mathletters}\label{octetFrags}
\begin{eqnarray}
D_{q\to\psi}(\mu^2, z) &=& \frac{2 \alpha_s^2}{9 M_\psi^3}
\,\langle O_8^\psi(^3S_1)\rangle\, \left\{\frac{(z-1)^2+1}{z}
\log\left[\frac{\mu^2}{M_\psi^2(1-z)}\right] - z\right\},\\
D_{g\to\psi}(\mu^2,z) &=& \frac{\pi\alpha_s}{3M_\psi^3}
\,\langle O_8^\psi(^3S_1)\rangle\, \delta(1-z),
\end{eqnarray}
\end{mathletters}
where
\begin{equation}
 O_8^\psi(^3S_1)=\chi^\dagger \sigma_iT^a\psi(a^\dagger_\psi a_\psi)
\psi^\dagger \sigma_i T^a \chi.
\end{equation}
$T^a$ is a color generator, while $\chi$ and $\psi$ are two 
component NRQCD spinors.  Inserting these fragmentation
functions into Eq.~(\ref{fragForm}) and matching onto the QCD
calculation, Eq.~(\ref{fragRate}), we obtain the coefficient functions
$C_q$ and $C_g$
\begin{mathletters}\label{WilsonCoeff}
\begin{eqnarray}
\label{WilsonCoeffa}
C_q(\mu^2,z) &=& \Gamma(Z\to q\bar q)\, \delta(1-z),\\
C_g(\mu^2,z) &=& \frac{4\alpha_s}{3\pi}\, \Gamma(Z\to q\bar q)
\left\{\frac{(z-1)^2+1}{z}\log\left[\frac{(1-z)z^2M_Z^2}{\mu^2}\right]
-z\right\}.
\end{eqnarray}
\end{mathletters}%
Note that there is no physical distinction between what we call gluon
fragmentation and what we call quark fragmentation, as it is always
possible to shift some finite piece from one to the other.  However,
factorization dictates that we match in such a way as to make the
Wilson coefficients independent of the long distance physics,
i.e. $M_\psi$.

Once we choose the scale $\mu$ to be $O(M_Z)$, there are no large logs
in the Wilson coefficients. They have been shuffled into the
fragmentation functions and can be resummed via the DGLAP
\cite{DGLAP} equations.  All the leading logs, of order $O[\alpha_s^2
\log(M_Z^2/M_\psi^2)]$, will reside in the quark fragmentation
function. The contribution from gluon fragmentation is then
subleading, contributing only at $O(\alpha_s^2)$ to the rate.  We
choose to keep this contribution in order to reduce the $\mu$
dependence of our result, although we do not claim accuracy to the
level $\alpha_s^2$. There are other terms of this order, arising from
both two-loop running and the $\alpha_s^2$ corrections to the initial
gluon fragmentation function, that have not been included.

In addition to the color octet contribution 
we include the contribution from
the color singlet, which is formally $O(\alpha_s^2 v^3)$, but as mentioned
above is numerically smaller than the octet contribution. 
Again using the Collins-Soper definition for the singlet 
fragmentation function we find

\begin{equation}\label{singFrag}
D^{(1)}_{c\to\psi}(\mu^2, z) = \frac{128\alpha_s^2}{243 M_\psi^3}\, \langle
O_1^\psi(^3S_1)\rangle\,\frac{z(1-z)^2}{(2-z)^6}(16-32z+72z^2-32z^3+5z^4),
\end{equation}
which agrees with \cite{BCY,MA}.

\section{$\lowercase{z}\sim 1$ Resummation}

The resummation of the $\log(M_\psi^2/M_z^2)$ is accomplished via the
usual renormalization group analysis of the fragmentation functions.
The evolution equations are given by
\begin{mathletters}\label{evolution}
\begin{eqnarray}
\mu\frac{dD_q(\mu^2,z)}{d\mu} &=& \frac{\alpha_s(\mu^2)}{\pi}
\left\{P_{q\to qg}\ast D_q(\mu^2) + 
P_{q\to gq}\ast D_g(\mu^2)\right\},\\
\mu\frac{dD_g(\mu^2,z)}{d\mu} &=& \frac{\alpha_s(\mu^2)}{\pi}\left\{
\sum_{j=1}^{2 n_f}
P_{g\to q\bar q}\ast D_{q}(\mu^2) + 
P_{g\to gg}\ast D_g(\mu^2)\right\},
\end{eqnarray}
\end{mathletters}%
where the functions $P$ are the standard splitting functions.
We solved these equations numerically
using $m_c = 1.48 \GeV$, $\mu =  M_Z$, $\alpha_s(M_Z) = 0.118$, and
chose $\alpha_s(M_\psi)$ to be consistent with one-loop running from $M_Z$.

\begin{figure}[t]
\centerline{\epsfysize=11truecm  \epsfbox{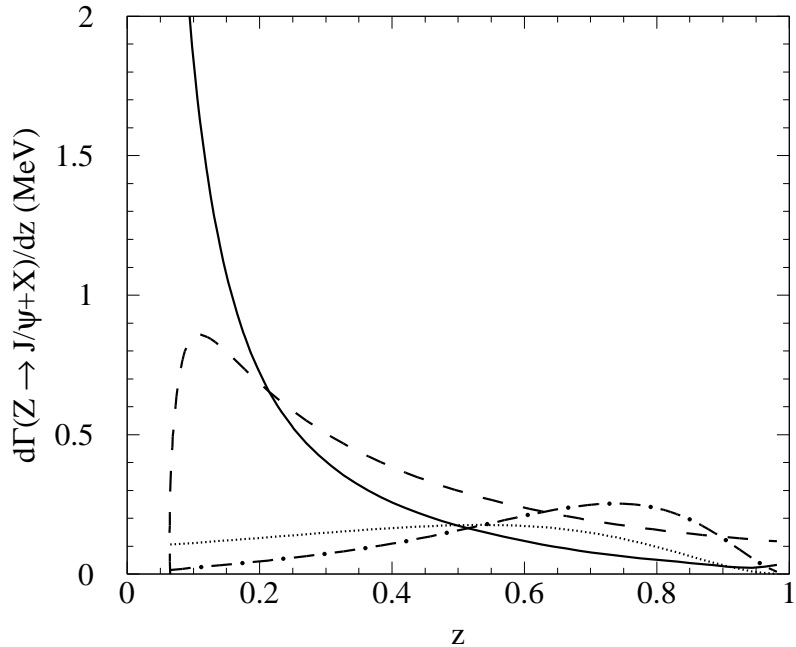} }
\tighten{
\caption[]{\it Differential rate 
${d\Gamma \over dz} $ for the octet channel with (solid line) and
without (dashed line) evolution, and for the singlet channel with
(dotted line) and without (dot-dashed line) evolution, as a function
of $z = 2 E_\psi/ M_Z$.  The octet matrix element has been extracted
from the Tevatron (see text).  }}
\end{figure}

The contributions to the differential rate from both the evolved octet
(solid line) and evolved singlet (dotted line) fragmentation functions
are displayed in Fig.~1, in units of MeV.  We used the Tevatron extraction
\cite{ChoLeib} to obtain the normalization of the octet fragmentation
functions.  For comparison, the octet contribution without evolution
is shown in dashed lines, and the singlet contribution without
evolution in dot-dashed lines.  Resummation of the
$\log(M_\psi^2/M_Z^2)$ terms greatly enhances both the octet and
singlet rates at small $z$. However, as mentioned above we should not
trust this result in the small $z$ regime.

At this point we should point out that there are some values of $z\sim
1$ where the differential rate is not trustworthy. When $z$ approaches
within $v^2$ of one, we begin to probe the hadronic structure of the
quarkonium state, and the expansion in $v$ breaks down\cite{IZR,MB}.
To correctly describe quarkonium production at the edge of phase space
we must introduce a structure function \cite{RW,BRW} which resums all
the large non-perturbative corrections. Indeed, LEP would be an ideal
place to study these structure functions if there were more data
available. In the total cross section however, or any sufficiently
smeared version of the differential rate, such as the first ten or so
moments , the expansion in $v$ is well behaved.

\section {Resummation for small $\lowercase{z}$}

The lower limit on $z$ is $z_{\rm min} = 2 M_\psi/M_Z$, so that when
$z\sim z_{\rm min}$, we need to include a resummation of the $\log(z)$ 
terms.  Indeed, the generic term in the decay rate
has pieces of the form $\alpha_s^{n+1} \log^l(z^2)
\log^m(M_Z^2/M_\psi^2),~ (l+m=2n-1)$.  We thus treat $\log(z^2)$ to be
of the same order as $\log(M_\psi^2/M_Z^2)$, and resum all terms of
the above form. This problem was first encountered in the calculation
of jet multiplicities, where it was noticed that the splitting
functions are highly singular at small $z$ and need to be resummed.
The results of these calculations led to predictions for the shape of
the hadron multiplicties (under some assumptions of quark-hadron
duality) which fit the data extremely well\cite{opaldata}. Indeed, the
results make the striking prediction that there should be a
suppression at small $z$ which sets in at higher $z$ than would be
expected from just phase space suppression. This suppression is due to
angular ordering of soft gluon emission and is a consequence of gluon
coherence which is naively missed in canonical ladder
resummations\cite{Mul0}.

Our calculation differs from those previous calculations of hadron
multiplicities in that we can actually calculate the hadronic
fragmentation function in a systematic fashion, in terms of some (in
our case effectively one) unknown matrix element. Thus, we are
concerned with the normalization as well as the shape of the
differential decay rate. There are several different formalisms for
handling the coherent gluon emission problem\cite{rus,italy,MulI}.  We
choose to follow the formalism developed by Mueller \cite{MulI,MulII}, and
we refer the reader to these papers for details.

\begin{figure}[t]
\centerline{ \epsfbox{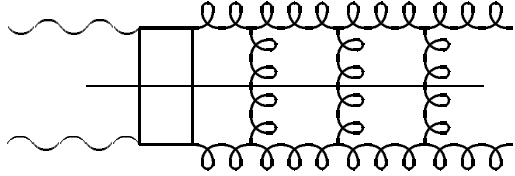} }
\tighten{
\caption[]{\it Typical ladder diagrams which lead to the leading logs.
To include coherence effects, angular ordering is imposed such that
the angle between two branching partons is smaller than the angle of
the previous pair.}}
\end{figure}

We begin by noticing that it is the gluon splitting function which is
most singular at small z, $P_{g\to gg}(z) \approx 2 C_A/z$. Thus, at
leading order as defined above, the branching will all come
from gluon splitting once the initial quarks mix into a gluon. For the
moment let us consider pure gluon splitting.  All the leading logs
come from ladder diagrams of the form shown in Fig.~2 (without the initial
quark box).  The coherence
issues can be skirted by imposing  angular ordering on the gluons
such that the angle between two branching partons is smaller than the
angle of the previous pair\cite{bcmm}.  This allows us to rewrite the series via
the integral equation
\begin{equation}\label{smallzevolve}
zD_g(\hat t,z) = \delta(1-z) + 
  \frac{\alpha_s C_A}{\pi} \int_z^1\frac{dz'}{z'}
     \int_{\hat t}^1\frac{d\hat t'}{\hat t'}z'D_g(\hat t',z'),
\end{equation}
where $C_A = 3$ and at the end of the calculation $\hat t$ is taken to
be $\hat t=M_\psi^2/M_z^2 z^2$ in order to enforce the angular
ordering.  For now, $D_g(\hat t,z)$ has been normalized to 1.
Iterating this equation leads to
\begin{equation}\label{iterate}
D_g(M_Z^2/M_\psi^2,z) = \frac{1}{z} 
  \sum_{m=1}^\infty \left(\frac{\alpha_sC_A}{\pi}\right)^m 
  \frac1{m!} \log^m\left(\frac{M_Z^2z^2}{M_\psi^2}\right) 
  \frac1{(m-1)!} \log^{m-1}\left(\frac1z\right). 
\end{equation}
Taking moments of Eq.~(\ref{iterate}) gives
\begin{eqnarray}
D_{gn}(M_Z^2/M_\psi^2) &=& \int_0^1 dz z^{n-1} D_g(M_Z^2/M_\psi^2,z) 
   \nonumber \\
&=& 1 + \sum_{m=1}^\infty\left(\frac{\alpha_s C_A}{\pi}\right)^m
    \sum_{k=0}^m\frac{(-2)^k(m+k-1)!}{(m-1)!k!(m-k)!}
    \frac{\log^{m-k}(M_Z^2/M_\psi^2)}{(n-1)^{k+m}}.
\end{eqnarray}
The above sums have a simple closed form. After  reinserting the 
normalization
\begin{equation}
D_{gn}(M_Z^2/M_\psi^2) = \frac{\pi \alpha_s}{3 M_\psi^3}
   \,\langle O_8^\psi(^3S_1)\rangle\,C_n e^{\gamma_n \log(M_Z^2/M_\psi^2)},
\end{equation}
where
\begin{eqnarray}
\gamma_n &=& \frac14\left[\sqrt{(n-1)^2 + 8\alpha_sC_A/\pi}-(n-1)\right],\\
C_n &=& 1 - \frac{2 \gamma_n}{\sqrt{(n-1)^2 + 8\alpha_sC_A/\pi}}.
\end{eqnarray}
$\gamma_n$ is the resummed diagonal anomalous dimension of the
fragmentation function. If we expand in $\alpha_s$, then it
correctly reproduces the most singular pieces of the previously
calculated two-loop splitting function\cite{2loop}. 

At small $z$, the leading contribution to the quark fragmentation
function comes from the gluon fragmentation function above, convoluted
with $P_{q\to gq}(z) \approx 2 C_F/z$.  Summing up all ladder diagrams
of the form shown in Fig.~2 leads to
\begin{eqnarray}\label{smallxquark}
D_q(M_Z^2/M_\psi^2,z) &=& \frac{\pi\alpha_s}{3 M_\psi^3}
   \,\langle O_8^\psi(^3S_1)\rangle\,\frac{C_F}{C_A} \nonumber\\
&&  \times \frac{1}{z}\sum_{m=1}^\infty \left(\frac{\alpha_sC_A}{\pi}\right)^m 
  \frac1{m!} \log^m\left(\frac{M_Z^2z^2}{M_\psi^2}\right) 
  \frac1{(m-1)!} \log^{m-1}\left(\frac1z\right),
\end{eqnarray}
with moments
\begin{equation}\label{Dqmoments}
D_{qn}(M_Z^2/M_\psi^2) = \frac{\pi \alpha_s}{3 M_\psi^3}
   \,\langle O_8^\psi(^3S_1)\rangle\,\frac{C_F}{C_A}
   C_n e^{\gamma_n \log(M_Z^2/M_\psi^2)}.
\end{equation}
Note that expanding Eq.~(\ref{smallxquark}) to leading order and
convoluting with twice the coefficient function,
Eq.~(\ref{WilsonCoeffa}) (for quark and antiquark fragmentation),
recovers the tree-level differential rate, Eq.~(\ref{fragRate}), in
the small $z$ limit.

We can now use the renormalization group to improve
Eq.~(\ref{Dqmoments}).  After running we are left with, 
\begin{equation}
D_{qn}(M_Z^2/M_\psi^2) = \frac{\pi \alpha_s}{3 M_\psi^3}
   \,\langle O_8^\psi(^3S_1)\rangle\,\frac{C_F}{C_A}\,
   C_n(\alpha_s(M_Z^2))\,\exp\left(\int_{M_\psi^2}^{M_Z^2}
   \frac{d\mu^2}{\mu^2} \gamma_n(\alpha_s(\mu^2))\right).
\end{equation}
This equation can now be inverted back to $z$ space leading to the result
%
%\begin{equation}
%D_q(Q^2,z) = \frac1{2\pi i}
%   \int_{C-i\infty}^{C+i\infty} dn x^{-n}D_{qn}(Q^2).
%\end{equation}
%

%
\begin{equation}
D_q(M_Z^2,z) \approx \frac{\pi \alpha_s}{3 M_\psi^3}
   \,\langle O_8^\psi(^3S_1)\rangle\,\frac{C_F}{C_A}\,
   C_{n_0}(\alpha_s(M_Z^2))\frac1{2z\sqrt{\pi a}}
   \exp\left[c -\frac1{4a}\left(\log\frac1z - b\right)^2\right],
\end{equation}
where
\begin{eqnarray}
n_0 &=& 1-\frac1{2a}\left(\log\frac1z-b\right), \\
a &=&\frac1{48b_0}\left(\sqrt{\frac{2\pi}{C_A\alpha_s(M_Z^2)^3}} - 
	\sqrt{\frac{2\pi}{C_A\alpha_s(M_\psi^2)^3}}\right), \\
b &=& \frac1{4b_0\alpha_s(M_Z^2)} - \frac1{4b_0\alpha_s(M_\psi^2)}, \\
c &=& \frac1{b_0}\left(\sqrt{\frac{2C_A}{\pi\alpha_s(M_Z^2)}} - 
	\sqrt{\frac{2C_A}{\pi\alpha_s(M_\psi^2)}}\right),
\end{eqnarray}
and $b_0$ is the coefficient of the one-loop beta function.  This
result was reached in the saddle point approximation where $1-n_0$ is
small and thus should not be trusted for $z$ values larger than
$z\sim0.2$. Subleading corrections to this result can be systematically
included by properly adapting the formalism discussed in
\cite{MulII,web}, where they were found to be of order 
$\sqrt{\alpha_s}$ at the peak\footnote{We have checked using 
the results of \cite{MulII}, that the numerical value of the corrections
to the coefficient function are actually quite a bit smaller than this.}.
We should note that we expect the relative size of the
subleading corrections to the total rate to be smaller than those  
for the differential rate, given that in the differential
rate the subleading terms are down by $\log(z^2)$, whereas in the
total rate they are down by $\log(M_Z^2/M_\psi^2)$.

\section {Extraction of the Matrix Element}
We now have leading log expressions for the total and differential
rates in the small and large $z$ regions. Before we interpolate between
the two we need to determine their respective regions of validity.
Let us investigate the size of the contributions we are 
neglecting in the large $z$ region.  The
generic term in the decay rate has the form
$\alpha_s^{n+1}\log^l(z^2)\log^m(M_Z^2/M_\psi^2)$, where $l+m = 2n-1$.
The first term that is not included in the large $z$ resummation is
suppressed by $\log(z)/\log(M_\psi/M_Z)$.  Thus a conservative
value for the lower value of $z$ is $ \sim 0.5$.

The small $z$ resummation is not valid at large $z$ for two reasons.
First, we have used the saddle point approximation to compute the
inverse Mellon transform, which is only valid for $n_0-1$ small.
Second, we have neglected less singular terms in computing the small
$z$ fragmentation result. Therefore, we will trust the small $z$
result only up to  the peak, where $n_0-1 \approx 0$, but not for larger $z$.
We believe this to be a conservative upper bound on $z$ for this
approximation.

Thus, to obtain the full differential cross section, we interpolate
between the small $z \lesssim 0.2$ and large $z\gtrsim 0.5$ resummations
using the data in this region as our guide.  We checked that varying
the interpolation while still staying within the error bars, changes
the result by at most $25\%$.  Furthermore, we tested the sensitivity
of our predictions to increasing and decreasing the lower bound on $z$
in the large $z$ regions as well as increasing the upper bound on the
small $z$ region. These variations do not appreciably change the total
rate or the first moment.  Note that {\it decreasing} upper bound on
small $z$ regions does drastically change the results.  However, taking the
maximum value of $z$ to be below $z_{peak}$ is not a reasonable thing to
do however, given that the peak is the position where the saddle point
approximation is trustworthy. In addition, the data for hadron
multiplicites fits the resummed predictions extremely well near the
peak\cite{opaldata}.

In our final result we also included the  non-fragmentation corrections
given by the difference between Eqs.~(\ref{exact}) and
(\ref{fragForm}). These are significant at very small $z$, and
contribute to the total rate even when $M_\psi/M_Z \to 0$.

Since the data includes feed-down from excited charmonium states, the
rate should be written in terms of the effective matrix elements
\cite{ChoLeib}
\begin{eqnarray}\label{Oeff}
\langle \hat O_8^{\psi(n)}(^3S_1)\rangle &\equiv& \sum_{m \ge n}
\langle O_8^{\psi(m)}(^3S_1)\rangle\ {\rm BR}(\psi(m)\to \psi(n) +X),
\nonumber \\ 
\langle \hat O_1^{\psi(n)}(^3S_1)\rangle &\equiv& \sum_{m \ge n}
\langle O_1^{\psi(m)}(^3S_1)\rangle\ {\rm BR}(\psi(m)\to \psi(n) +X).
\end{eqnarray}
Saturating the excited states with the $\psi'$ and $\chi_{cJ}$, and
using values for $\langle O_8^{\psi}(^3S_1)\rangle$, $\langle
O_8^{\psi'}(^3S_1)\rangle$, and $\langle O_8^{\chi_J}(^3S_1)\rangle$
from \cite{ChoLeib} gives $ \langle \hat O_8^{\psi}(^3S_1)\rangle =
0.014 \pm 0.002 \GeV^3.$ This number has at least a factor of two
theoretical uncertainty.  Extracting the analogous color singlet
matrix elements from the $\psi$ and $\psi'$ electronic widths
\cite{BBL} gives $ \langle \hat O_1^{\psi}(^3S_1)\rangle = 1.45 \pm
0.10 \GeV^3$.

Combining the singlet and octet fragmentation contributions gives the
total differential rate.  Integrating over $\psi$ energies yields a
total branching ratio of
\begin{equation}\label{BranchingRatios}
\rm{ BR}(Z \to {\rm prompt}\ J/\psi + \rm{X} ) =
 \Biggl( 1.47\  {\langle \hat O_8^{\psi}(^3S_1)\rangle \over 0.014 \GeV^3 }
+  0.47\ {\langle \hat O_1^{\psi}(^3S_1)\rangle \over 1.45 \GeV^3 } 
\Biggr) \times\ 10^{-4} ,
\end{equation}
compared to a branching ratio of 
$(1.93\ \langle \hat O_8^{\psi}(^3S_1)\rangle /0.014 \GeV^3 
+ 0.68\ \langle \hat O_1^{\psi}(^3S_1)\rangle /1.45 \GeV^3 
 ) \times\ 10^{-4} $ from \cite{Cho}. 
%The tree-level  rates would be decreased by $28\%$ if \cite{Cho} 
%obtained $\alpha_s(M_\psi)$ by running from $M_Z$, but there is no
%justification for this unless the large logarithms are resummed. 

The total branching ratio has been measured to be \cite{expt,ALEPH}
\begin{eqnarray}\label{Opal}
{\rm BR}(Z \to {\rm prompt}\ J/\psi +X) &=& 
 (1.9 \pm 0.7 \pm 0.5 \pm 0.5) \times 10^{-4}\ {\bf OPAL} 
\nonumber \\
{\rm BR}(Z \to {\rm prompt}\ J/\psi +X) &=& 
 (3.0 \pm 0.8 \pm 0.3 \pm 0.15 ) \times 10^{-4}\ {\bf ALEPH}
\nonumber \\
{\rm BR}(Z \to {\rm prompt}\ J/\psi +X) &=&
 (2.7 \pm 1.2 ) \times 10^{-4}\ {\bf L3}
\nonumber \\
{\rm BR}(Z \to {\rm prompt}\ J/\psi +X) &=&
 (4.4^{+3.6}_{-3.0} ) \times 10^{-4}\ {\bf DELPHI}\ .
\end{eqnarray}
For OPAL and ALEPH, the uncertainties from left to right
are statistical, systematic, and model-dependent, while for
L3 and DELPHI, they are purely statistical. 
We use the LEP average to extract a value 
for the effective octet matrix element of 
\begin{equation}\label{Oeight}
\langle \hat O_8^{\psi}(^3S_1)\rangle = (0.019 \pm 0.005_{\rm stat} 
\pm 0.010_{\rm theo})\GeV^3,
\end{equation}
where the first uncertainty is purely statistical, and the second
is theoretical. The theoretical uncertainty comes from adding in
quadrature roughly $30\%$ contributions from perturbative corrections 
suppressed by $\alpha_s(M_\psi)$, higher order matrix elements 
suppressed by $v^2$, and subleading logs.  

Since the LEP experiments include feed-down from $\psi'$ and $\chi_J$,
we cannot directly compare our extraction of $\langle \hat
O_8^{\psi}(^3S_1)\rangle$ with those of $ \langle
O_8^{\psi}(^3S_1)\rangle $ in \cite{Ocl,Okk,Obk}, but our value of
$\langle \hat O_8^{\psi}(^3S_1)\rangle$ is comparable with the central
value from \cite{ChoLeib}.  While our statistical uncertainties are
larger, our theoretical uncertainties are under good control. Since
this cannot be said of Tevatron extractions, where theoretical
uncertainties dominate, we believe Eq.~(\ref{Oeight}) represents the
most reliable extraction currently available.

\begin{figure}[t]
\centerline{\epsfysize=11truecm  \epsfbox{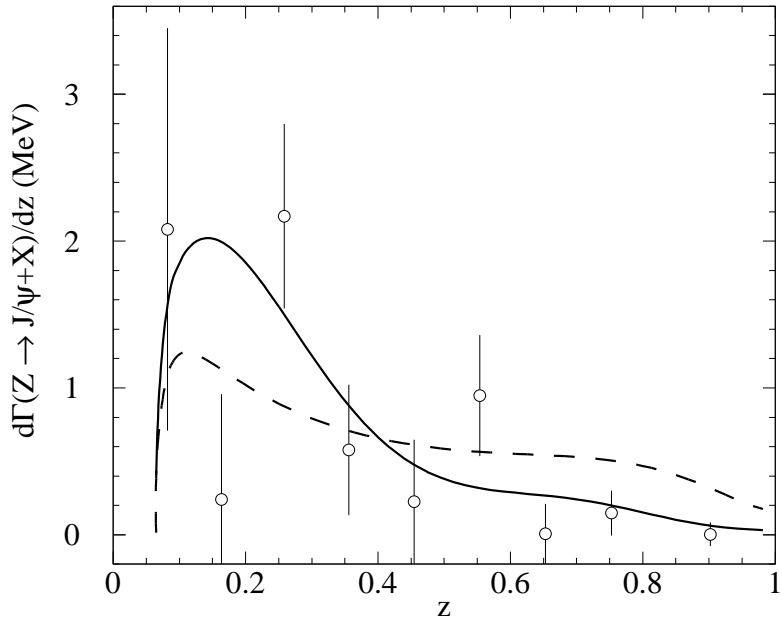} }
\tighten{
\caption[]{\it Differential rate ${d\Gamma \over dz} $ as a function of
$z = 2 E_\psi/ M_Z$ vs data.  The dashed line is the sum of the tree-level
octet and singlet results and the solid line is the interpolation
between the large and small $z$ region octet resummation plus the
singlet resummation.}}
\end{figure}

Fig.~3 shows the complete differential rate using our extracted value
for $\langle \hat O_8^{\psi}(^3S_1)\rangle$.  The solid line is the
sum of the resummed singlet fragmentation and the interpolation
between the large and small $z$ regions for octet fragmentation.  The
dashed line is the tree-level result and the data is from the ALEPH
collaboration \cite{ALEPH} with efficiency correction from
\cite{Rousseau}.  Given the large errors, it is difficult to make any
definite statements, but it seems the resummed rate fits the data
better at this time.  The effect of evolution is to enhance the
small-$z$ peak, a promising experimental signature of the octet
component.  A manifestation of this signature is a relatively small
first moment, which we find to be
\begin{equation}
\frac1{\Gamma(Z \to {\rm prompt}\ J/\psi + \rm{X} ) } \int 
\frac{d\Gamma(Z \to {\rm prompt}\ J/\psi + \rm{X} )}{dz}\,z\,dz = 0.30,
\end{equation}
while the tree-level differential rate, Eq.~(\ref{exact}), gives $\sim
0.5$.  A very rough estimate of this quantity obtained from the data
\cite{ALEPH} suggests a value of $0.26 \pm 0.10$.  This is in sharp
contrast to the color singlet prediction.  The tree-level color
singlet decay rate predicts the ratio of the first moment over the
zeroth moment to be $0.62$.  Resummation softens the color singlet
decay rate, but the ratio is still too large, $0.47$.  The ratio is
independent of the color singlet matrix element.  Therefore, even if
the color singlet rate were arbitrarily increased so that the color
singlet model fit the experimental branching ratio, the ratio of
moments would not fit the data.  A rigorous extraction of the first
moment by the experimental groups could provide an extremely clean,
quantitative test of the NRQCD approach.

In conclusion, we have resummed large logarithms in the rate for
prompt $J/\psi$ production at LEP to obtain the leading-order
prediction.  We predict a branching ratio that is slightly smaller
than the tree-level prediction\cite{Cho,CKY}. Moreover, we have eliminated a
factor of 2 uncertainty in the tree-level result.  Matching our
branching ratio to LEP data yields an octet matrix element with
substantially smaller theoretical uncertainties than from hadronic
processes\cite{Ocl,Okk,ChoLeib,Obk}.  The differential decay rate is
dramatically softer than previous calculations.  The small-$z$ peak in
the differential distribution represents a clear signature of the
octet mechanism that we regard as strong motivation for continued
analyses by the LEP experimental groups.  A measurement of the first
moment would be a particularly interesting.

\acknowledgments 
We thank and Maneesh Wadwha and Tom Ferguson for discussions on the data. 
We are especially grateful to D. Rousseau who supplied us with the
energy distribution and  Fabrizio Odorici who supplied us with
the cut cross section.
This work was supported in part by the Department of Energy under
grant number DOE-ER-40682-143.

{\tighten

} %end tighten (references)


\begin{references}

\bibitem{BBL}
G. T. Bodwin, E. Braaten, and G. P. Lepage, Phys. Rev. D51 (1995)
1125; Erratum-ibid. D55 (1997) 5853.

\bibitem{BratFlem}
E. Braaten and S. Fleming, Phys. Rev. Lett. 74 (1995) 3372.

\bibitem{Ocl}
B. Cano-Coloma and M. A. Sanchis-Lozano, Nucl. Phys. B508 (1997) 753.

\bibitem{Okk}
B. A. Kniehl and G. Kramer, hep-ph/9803256 (unpublished).

\bibitem{ChoLeib}
P. Cho and A. K. Leibovich, Phys. Rev. D53 (1996) 150;
ibid. D53 (1996) 6203.

\bibitem{Obk}
M. Beneke and M. Kr\"amer, Phys. Rev. D55 (1997) 5269.

\bibitem{KS} J.H. K\"uhn and H. Schneider, Phys. Rev. D24 (1981) 2996;
Z. Phys. C11 (1981) 263.

\bibitem{BCK}V. Barger, K. Cheung and W.Y. Keung, Phys. Rev. D41 (1990) 1541.

\bibitem{BCY}E. Braaten, K Cheung and T.C. Yuan, Phys. Rev. D48 (1993) 4230.

\bibitem{Cho}
P. Cho, Phys. Lett. B236 (1996) 171.

\bibitem{CKY}
K. Cheung, W-Y Keung, and T. C. Yuan, Phys. Rev. Lett. 76 (1996) 877.

\bibitem{YQC} F. Yuan, C.F. Qiao, K.T. Chao, Phys. Rev. D57 (1998) 610.

\bibitem{rus}
Yu. L Dohkshitzer et. al., ``Basics of Pertrubative QCD'', 
Editions Frontiers, Gif-sur-Yvette (1991).

\bibitem{italy}A. Bassetto, M. Ciafaloni and G. Marchesini, Phys. Rep. 100 (
1983) 201.

\bibitem{CollinsSoper}
J. C. Collins and D. E. Soper, Nucl. Phys. B194 (1982) 445.

\bibitem{Fleming2}
S. Fleming, Phys. Rev. D50 (1994) 5808.

\bibitem{BY} E. Braaten and T.C. Yuan, Phys. Rev. Lett. 71 (1993) 1673.

\bibitem{MA}J.P. Ma Phys. Lett. B332 (1994) 398.

\bibitem{DGLAP}
V. Gribov and L.N. Lipatov, Sov. J. Nucl. Phys 15 (1972) 78.\\
Yu. L. Dokshitzer, JETP 73 (1977) 1216.\\
G. Altarelli and G. Parisi, Nucl. Phys. B126 (1977) 298.


\bibitem{IZR} I.Z. Rothstein, Int. J. Mod. Phys. A12 (1997) 3857.

\bibitem{MB} M. Beneke hep-ph/9703429, Lectures given at 24th Annual
SLAC Summer Institute.

\bibitem{RW} I.Z. Rothstein and M.B. Wise, Phys. Lett. B402 (1997) 346.

\bibitem{BRW}M. Beneke I.Z. Rothstein and M.B. Wise, Phys. Lett. B408 (1997)
373.

\bibitem{opaldata}M.Z. Akrawy  et. al. Phys. Lett. B247 (1990) 617. 

\bibitem{Mul0} A.H. Mueller, Phys. Lett B104 (1981) 161.

\bibitem{MulI}A.H. Mueller, Nuc. Phys. B213 (1983) 85.

\bibitem{MulII} A.H. Mueller, Nuc. Phys. B228 (1983) 351.
\bibitem{bcmm}A. Bassetto et. al., Nuc. Phys. B207 (1982) 189.



\bibitem{2loop}G. Curci, W. Furmanski and R. Petronzio, Nuc. Phys.
B175, (1980) 27. E.G. Floratos, C Kounnas and R Lacaze, Nuc. Phys.
B192 (1981) 417.



\bibitem{web}C.P. Fong and B.R. Webber, Nucl. Phys. B355 (1991) 54.



\bibitem{expt}
OPAL Collaboration, Phys. Lett B368 (1996) 343; \\ 
M. Wadhwa, L3 Collaboration, talk given at QCD97 Montpellier, 1997;\\
P. Abreu et al., DELPHI Collaboration, Phys. Lett. B341 (1994) 109.

\bibitem{ALEPH}
ALEPH Collaboration, submission to the 1997 EPS-HEP conference, No. 624.

\bibitem{Rousseau}
D. Rousseau, private communication.



\end{references}
\end{document}